\documentclass[aps,twocolumn,superscriptaddress,showpacs,amssymb,prb,floatfix,longbibliography]{revtex4-1}
\usepackage{graphicx }
\usepackage{dcolumn}
\usepackage{bm}
\usepackage{color}
\usepackage{amsmath}
\usepackage{amssymb}
\usepackage{ulem}

\usepackage[]{hyperref}
\hypersetup{colorlinks=true,linkcolor=blue,citecolor=blue,urlcolor=blue,pdfpagemode=UseNone}

\newcommand{\slrrtext}  {spin-lattice-relaxation rate}
\newcommand{\slrr}      {$T_1^{-1}$}

\newcommand{\niro}{Na$_2$IrO$_{3}$}

\begin{document}

\title{Impact of disorder on dynamics and ordering in the honeycomb lattice iridate Na$_2$IrO$_3$}

\author{R. Sarkar}
\affiliation{Institute of Solid State and Materials Physics, Technical University of Dresden, 01062 Dresden, Germany}
\author{Z. Mei}
\affiliation{Department of Physics, University of California, Davis, California
95616, USA}
\author{A. Ruiz}
\author{G. Lopez}
\affiliation{Department of Physics, University of California, Berkeley, California 94720}
\affiliation{Materials Science Division, Lawrence Berkeley National Laboratory, Berkeley, California 94720}
\author{H.-H. Klauss}
\affiliation{Institute of Solid State and Materials Physics, Technical University of Dresden, 01062 Dresden, Germany}
\author{J. G. Analytis}
\affiliation{Department of Physics, University of California, Berkeley, California 94720}
\affiliation{Materials Science Division, Lawrence Berkeley National Laboratory, Berkeley, California 94720}
\author{I. Kimchi}
\affiliation{JILA, NIST and Department of Physics, University of Colorado, Boulder, CO 80309}
\author{N. J. Curro}
\affiliation{Department of Physics, University of California, Davis, California
95616, USA}

\date{\today}
\begin{abstract}

Kitaev's honeycomb spin-liquid model and its proposed realization in materials such as $\alpha$-RuCl$_3$, Li$_2$IrO$_3$ and Na$_2$IrO$_3$ continue to present open questions about how the dynamics of a spin-liquid are modified in the presence of  non-Kitaev interactions as well as the presence of inhomogeneities. Here we use $^{23}$Na nuclear magnetic resonance to probe both static and dynamical magnetic properties in single crystal Na$_2$IrO$_3$. We find that the NMR shift follows the bulk susceptibility above 30 K but deviates from it below; moreover  below $T_N$ the spectra show a broad distribution of internal magnetic fields. Both of these results provide evidence for inequivalent magnetic sites at low temperature, suggesting inhomogeneities are important for the magnetism.  The spin lattice relaxation rate is isotropic and diverges at $T_N$, suggesting that the Kitaev cubic axes may control the critical quantum spin fluctuations. In the ordered state, we observe gapless excitations, which may arise from site substitution, emergent defects from milder disorder, or possibly be associated with nearby quantum paramagnetic states distinct from the Kitaev spin liquid.
\end{abstract}


\maketitle

In recent years there has been increasing interest in the so-called Kitaev materials  A$_2$IrO$_3$ (A=Na, Li), which are model systems for Kitaev honeycomb physics, similar to $\alpha$-RuCl$_3$ and Li$_2$RhO$_3$.\cite{KitaevMaterialsReview,Gegenwart2015,KimchiKitaevReview,JackeliKhaliullin,Takagi2019} The Ir has electronic configuration $5d^5$, and a combination of spin-orbit coupling, Coulomb interactions, and crystal field interactions give rise to a  Mott insulating state with a gap of 340\,meV.~\cite{CominNa2IrO3}  Importantly, the $j=1/2$ Ir spins in the honeycomb structure experience Ising interactions along different $x,y,z$-directions with the three neighboring spins in the lattice. These couplings are strongly frustrated, and theory predicts an exotic spin liquid ground state with itinerant, gapless Majorana fermion states.\cite{KimchiKitaevReview}  In addition to the Kitaev interaction, higher-order Heisenberg terms are relevant in \niro, giving rise to long-range zig-zag antiferromagnetic order of the Ir spins below 15 K. \cite{ZigZagNaIrO3,GegenwartNa2IrO3PRB2011,CaoNa2IrO3zigzag}
At high temperatures, the magnetic susceptibility exhibits Curie-Weiss behavior with an effective moment close to that expected for spin 1/2. \cite{KitaevA2IrO3PRL,SinghGegenwart}
Diffuse magnetic x-ray scattering experiments have provided compelling evidence for the presence of significant bond directional interactions, which suggest that the Kitaev interactions indeed dominate the magnetic degrees of freedom.\cite{HwanChun2015}

The low energy spin dynamics in \niro\ and their relation to the relative size of the Kitaev and Heisenberg interaction terms have remained unclear, however.  Complicating matters is the fact that disorder could potentially give rise to additional magnetic moments with their own low energy dynamics.  Nuclear Magnetic Resonance (NMR) is a powerful microscopic probe that can shed light on the low temperature behavior of the iridates. The NMR shift, $K$,  probes the intrinsic  spin susceptibility. Disorder and extrinsic effects can dominate the bulk magnetic response, precluding detailed understanding of the low temperature behavior.  Furthermore, the NMR \slrrtext, \slrr, probes the dynamical spin susceptibility, providing information about the low energy excitations that are present in the system.

NMR has played an important role in uncovering the physics of the related Kitaev honeycomb lattice material, $\alpha$-RuCl$_3$. In this system \slrr\ is strongly field- and temperature-dependent, reflecting the suppression of long-range order and emergence of a field-induced quantum spin liquid above 9 T.\cite{BaekRuCl3PRL} Both the bulk susceptibility, $\chi$, and \slrr\ are strongly anisotropic in this material. Whether or not the field-induced phase exhibits a spin-gap remains unclear, however.\cite{Jansa2018,ZhengRuCl3PRL} The spin-liquid phase of the related compound H$_3$LiIr$_2$O$_6$ has also been investigated by NMR.\cite{Kitagawa2018}  Li$_2$IrO$_3$ and \niro\ have been studied less. Large single crystals of  Li$_2$IrO$_3$ are difficult to grow, however a mosaic of several sub-mm crystals has been studied, revealing similar behavior to $\alpha$-RuCl$_3$.\cite{Li2IrO3NMR2019}  NMR and $\mu$SR studies of polycrystalline \niro\ have been reported recently which probe the phase diagram as a function of pressure and Li doping.\cite{Simutis2018}

Here we report detailed $^{23}$Na ($I=3/2$) NMR studies of a high quality single crystal of \niro, which reveal a broad static field distribution below 15 K, as well as a peak in \slrr\ associated with the critical dynamics of an antiferromagnetic transition.  In the paramagnetic state,  $K$ is temperature dependent and anisotropic, similar to the bulk susceptibility.  However, $K$ does not track $\chi$ over the entire temperature range, but deviates below a temperature $T^*\sim 30$ K.  Surprisingly, \slrr\ is isotropic, and in the ordered state the spin dynamics reveal no sign of the opening of a gap.  Rather, $(T_1T)^{-1}$ remains constant as $T\rightarrow 0$, suggesting that fluctuations of the Ir moments persist deep in the long-range ordered state and that  \niro\ is located in close proximity to a quantum spin liquid state.  The presence of disorder, possibly from Na-Ir site substitutions or the presence of stacking faults, provides a consistent explanation for these observations.

\begin{figure}
\includegraphics[width=\linewidth]{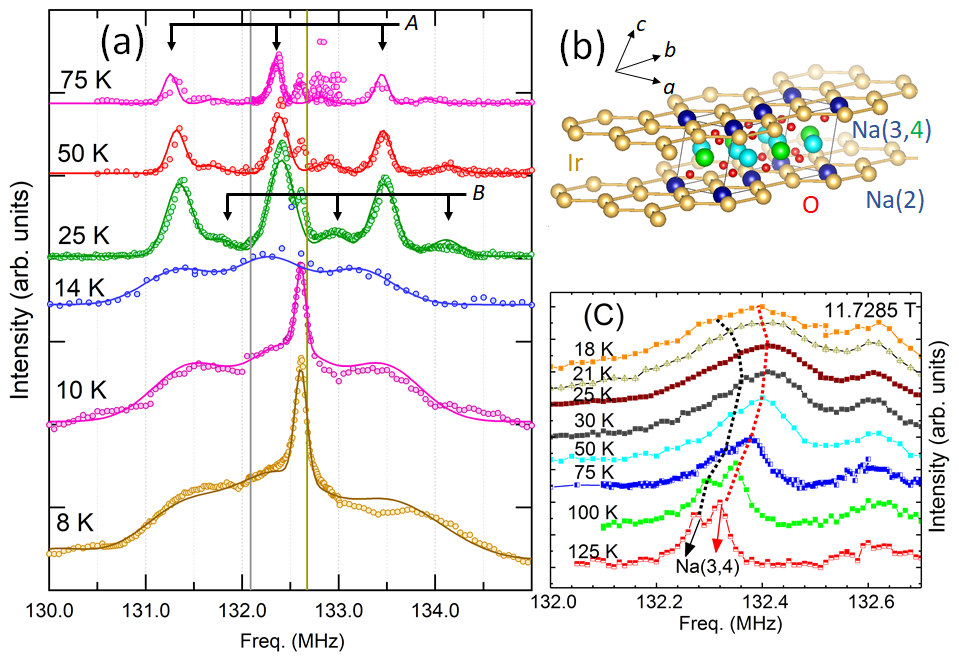}
\caption{\label{fig:spectra}
(a) $^{23}$Na spectra at fixed field $\mathbf{H}_0~||~c^*$ in \niro\ at several different temperatures. The vertical line at 132.09 MHz corresponds to the Larmor frequency of $^{23}$Na, and that at 132.67 MHz corresponds to the resonance frequency of metallic $^{63}$Cu. The solid lines are fits as described in the text. (b) Structure of \niro, indicating the three Na sites in the unit cell, and the honeycomb structure of the iridium atoms. Note that $c^*$ is normal to the planes, whereas $c$ is not.   (c) Detailed spectra of the Na(3) and Na(4) central transitions at several temperatures, showing the resolution of two separate peaks.}
\label{spectra}
\end{figure}

Single crystals of Na$_2$IrO$_3$ were prepared by mixing elemental Ir (99.9\% purity, BASF) with Na$_2$CO$_3$ (99.9999\% purity, Alfa-Aesar) in a 1:1.05 molar ratio. The mixture was ground for several minutes, then pressed into a pellet at approximately 3,000 psi. The pellet was then warmed in a furnace to 1050 $^{\circ}$C and held at temperature for 48 hours, before being cooled to 900 $^{\circ}$C over 24 hours and then furnace-cooled. Single crystals more than one square millimeter were then collected from the surface of the pellet. A  crystal of dimensions  1.1\,mm $\times$ 0.9\,mm $\times$ 0.07\,mm was selected and oriented  with magnetic field $H_0 = 11.73$ T applied parallel and perpendicular to the $c^*$ direction (normal to the $ab$ plane).   NMR experiments were performed  at a fixed field  for temperatures between 4K and 300 K. $^{23}$Na ($I=3/2$, $\gamma$\,=\,11.2625\,MHz/T, 100\% abundance) NMR spectra were collected by a home-built auto-tuning and matching NMR probe over broad frequency ranges. Silver wire was used for the NMR coil to avoid overlap between the $^{63}$Cu and $^{23}$Na resonances. 
$^{23}$Na spectra were acquired by collecting spin echoes  as a function of frequency, and \slrrtext\  measurements were conducted by observing the spin echo following an  inversion pulse at the central transition.

\begin{figure}[h]
\begin{center}
\includegraphics[width=\linewidth]{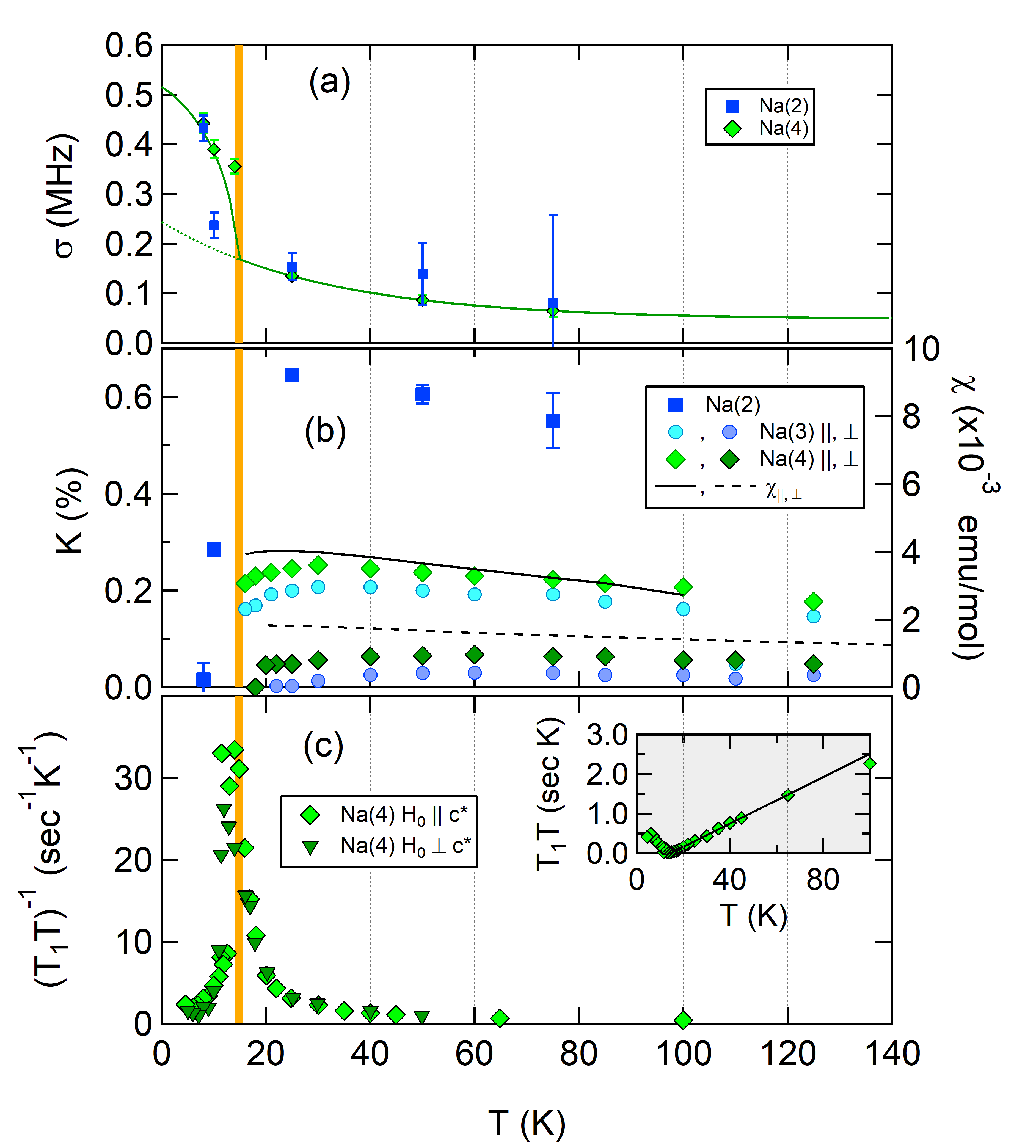}
\caption{ (a) Second moments of the Gaussian spectral functions used to fit the data shown in Fig. \ref{fig:spectra} versus temperature.  The solid and dashed lines are fits as described in the text. (b) NMR shifts of three sites and the bulk magnetic susceptibility (solid line) versus temperature. (c) $(T_1T)^{-1}$ measured at the Na(3) and Na(4) site as a function of temperature, for both $\mathbf{H}_0 ~ || ~c^*$ and $\mathbf{H}_0\perp c^*$.  The solid vertical orange line indicates $T_N$. (INSET) $T_1T$ versus $T$ for $\mathbf{H}_0 ~ || ~c^*$, with a linear fit (solid line), indicating a divergence at $T_N = 14.3\pm 0.1$ K.  }
\label{fig:parameters}
\end{center}
\end{figure}

Figure\,\ref{spectra} shows  $^{23}$Na NMR spectra collected at several representative temperatures. There are three nonequivalent Na sites in \niro\ (see Fig.~\ref{fig:spectra}(b)), each described by the Hamiltonian: $\mathcal{H} = \gamma \hbar \mathbf{\hat{I}}\cdot(1 + \mathbf{K})\cdot\mathbf{H}_0+ \frac{h\nu_{zz}}{6}\left[3\hat{I}_z^2 - \hat{I}^2 + \eta(\hat{I}_x^2 - \hat{I}_y^2)\right]$, where $\eta =(\nu_{xx} - \nu_{yy})/\nu_{zz}$, and $\nu_{\alpha\alpha}$ are the eigenvalues of the electric field gradient (EFG) tensor, and $\mathbf{K}$ is the NMR shift tensor.\cite{CPSbook} $\mathbf{H}_0$ is not necessarily parallel to any of the principle directions of either the NMR shift or EFG  tensors.   At high temperatures the spectra are considerably narrow and the satellite structure is clearly evident. There is also a temperature-independent background resonance at 132.67 MHz from metallic Cu.
Below 50K, a second set of resonances emerges, and below 14K the spectra become significantly broader, with one narrow peak centered at 132.6 MHz.  The narrow peak is temperature-independent, and we ascribe this to spurious background signal from $^{63}$Cu in the probe. The broad spectra arise from the $^{23}$Na in the crystal, which experience a range of internal magnetic fields in the magnetically ordered state.

We fit the spectra in Fig. \ref{fig:spectra}(a) to extract the NMR shifts, $K$, the quadrupolar splittings, $\nu_{c^*c^*}$, and the second moment, $\sigma$, for two $^{23}$Na sites, tentatively identified as sites A and B. The fitting was performed by exact diagonalization and accounts for the three quadrupolar-split resonances for both sites simultaneously.  Site A exhibits a higher intensity  with quadrupolar splitting $\nu_{c^*c^*}(A) = 1.056\pm 0.003$ MHz and NMR shift $K(A) = 0.242\pm 0.001\%$.  The spectrum for site B is approximately 1/3 in intensity, with a larger $\nu_{c^*c^*}(B) = 1.16\pm 0.02$ MHz and larger NMR shift $K(B) = 0.65\pm 0.01 \%$.  At lower temperatures, the central transition for site A splits into two separate peaks with slightly different NMR shifts, as shown in Fig. \ref{spectra}(c). We therefore identify site A as the Na(3) and Na(4) sites, located between the Ir layers, and site B as the Na(2), located within the Ir planes.  This assignment is supported by point charge calculations of the EFG for the three sites, which indicate a slightly larger EFG at the Na(2) site. The EFG for the two interplanar sites, Na(3) and Na(4) is similar, but they appear to have slightly different NMR shifts.  We are unable to determine which site corresponds to which shift, but for concreteness we assign Na(3) to the lower NMR shift.  At higher temperature, the shifts and quadrupolar splittings for these two sites are not distinct enough to resolve.  We find that the EFG parameters change by less than 5\% with temperature in the paramagnetic phase.

Figures \ref{fig:parameters}(a) and (b) displays the temperature dependence of the second moment and NMR shift versus temperature for both sites.  The linewidth increases with decreasing temperature, but increases by a factor of three at the Na(2) site below 14 K.  This increase reflects the presence of local internal magnetic fields present at the Na(2) site due to the onset of static magnetic order. The fact that the spectra do not reveal any sharp peaks below $T_N$ but rather are broad and featureless indicate a distribution of internal fields.  Similar results have been also observed in the ordered phase of RuCl$_3$.\cite{BaekRuCl3PRL} This observation suggests incommensurate magnetic order, however neutron and x-ray scattering studies indicate a commensurate zig-zag antiferromagnetic order,\cite{ZigZagNaIrO3,CaoNa2IrO3zigzag}  and muon time spectra exhibit oscillations indicating a well defined static internal field at the muon site, rather than a broad distribution of fields as expected for incommensurate order.\cite{Na2IrO3muSR} A possible explanation for this discrepancy is substitutional disorder among the Na and Ir lattice sites, as discussed below. Note that the Na is likely coupled to several nearest-neighbor Ir spins through a complex set of hyperfine couplings.  As a result, even a small level of disorder can quickly lead to a broad distribution of static hyperfine fields, both in terms of magnitude and direction.\cite{Dioguardi2010}   In the paramagnetic state, we find that the temperature dependence of the second moment can be fit empirically to $\sigma_1(T) = A + Be^{-T/T_0}$, with $A = 0.048$ MHz, $B = 0.196$ MHz, and $T_0 = 30.0$ K.  Below $T_N$ we include an extra mean-field broadening term, $\sigma(T) = \sigma_1(T) + \sigma_2(T)$, where $\sigma_2(T) = \sigma_0\sqrt{1-(T/T_N)^2}$, with $\sigma_0 = 0.27$ MHz, and $T_N = 14.3$ K.  This fit is shown as a solid line in Fig. \ref{fig:parameters}(a).

\begin{figure}[h]
\begin{center}
\includegraphics[width=\linewidth]{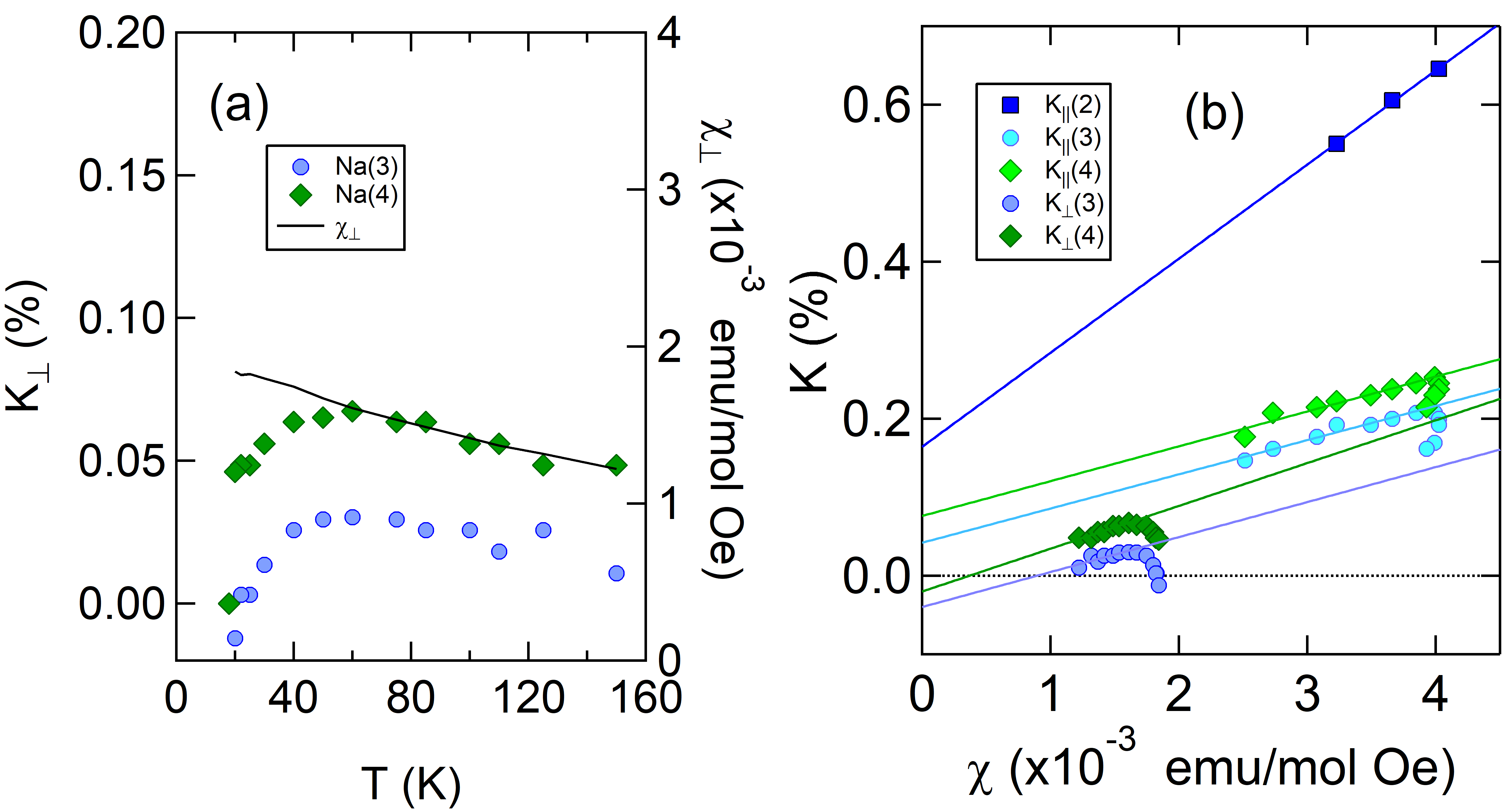}
\caption{(a) NMR shifts and magnetic susceptibility (solid line) versus temperature for the field oriented perpendicular to $c^*$. (b) NMR shifts versus susceptibility for both field directions. The solid lines are fits to the high temperature data, as described in the text, with fitting parameters detailed in Table \ref{tab:hyperfinepars}.}
\label{fig:shift}
\end{center}
\end{figure}

\begin{table}[htbp]
  \centering
  \caption{\label{tab:hyperfinepars} Hyperfine parameters from fits to the high temperature data as shown in Fig. \ref{fig:shift}.}
    \begin{ruledtabular}
    \begin{tabular}{lcccc}
    Site & $K_{||}^0$ (\%) &  $K_{\perp}^0$ (\%) &  $A_{||}$ (kOe/$\mu_B$) &  $A_{\perp}$ (kOe/$\mu_B$)\\
    \hline
    Na(2) & 0.06$\pm 0.02$ &- &6.70$\pm 0.26$ & - \\
    Na(3) & 0.04$\pm0.01$ & -0.04$\pm 0.01$& $2.43\pm0.30$ & 2.49$\pm0.40$\\
    Na(4) & 0.08$\pm0.02$ & -0.02$\pm 0.01$ & $2.48\pm 0.50$ & 3.05$\pm0.38$\\
        \end{tabular}%
    \end{ruledtabular}%
\end{table}%

The NMR shift shown in Figs. \ref{fig:parameters}(b) and \ref{fig:shift}(a) is compared with the bulk magnetic susceptibility, $\chi$ measured for field both parallel and perpendicular to the $c^*$ direction.   For both directions, $K$ increases with decreasing temperature down to a maximum $\sim 30$ K, and then decreases below. The NMR shift derives from the hyperfine coupling between the $^{23}$Na nuclear spins and the Ir electron moments, and the NMR shift should be proportional to $\chi$ as:  $K = A\chi + K_0$, where $A$ is the hyperfine coupling constant and $K_0$ is a temperature-independent constant. As seen in Fig. \ref{fig:parameters}(b), $K$ and $\chi$  exhibits similar behavior at high temperature, but below this temperature $K$ and $\chi$ no longer track one another.  This anomalous behavior is clearly evident in Fig. \ref{fig:shift}(b), where $K$ is plotted versus $\chi$  for both field directions.  At high temperatures $K$ and $\chi$ are linearly proportional with hyperfine constants given in Table \ref{tab:hyperfinepars}.  These values are about an order of magnitude larger than the direct dipolar coupling between the Na nuclei and the Ir electron spins (on the order of 0.5 kOe/$\mu_B$), and are consistent with a transferred hyperfine coupling due to wavefunction overlap.  Given this value of the hyperfine field, we can estimate the magnitude of the ordered moments as $S_0 \approx \sigma_0/\gamma A \sim 0.1\mu_B$/Ir. This order of magnitude is consistent with neutron scattering measurements that indicate 0.22$\mu_B/$Ir.

The breakdown of the linear $K-\chi$ relationship at low temperature is puzzling. In heavy fermions,  an NMR shift anomaly usually reflects the onset of coherence,\cite{Curro2012,Curro2009} however \niro\ is insulating and there should be no such effect. It is possible that there are different hyperfine couplings to the orbital and spin moments of the Ir, but the spin and orbital susceptibilities will exhibit the same temperature dependence due to the strong spin-orbit coupling. In this case, the $K-\chi$ relationship will reflect a renormalized effective hyperfine coupling,\cite{NissonCEFSOC2016}  but will not exhibit an anomaly as we observe. Such an anomaly usually indicates the presence of multiple magnetically-active sites with different temperature dependences. Substitutional disorder, with some fraction of the Ir sites located at the Na sites rather than in the honeycomb lattice structure, could therefore explain this behavior.  A similar breakdown of the $K-\chi$ relationship has also been observed in RuCl$_3$.\cite{BaekRuCl3PRL}

In addition to the spectra we also measured the spin-lattice relaxation rate, \slrr\ at the central transition of the Na(3,4) sites for both field orientations. The magnetization recovery was fit to a stretched exponential appropriate for the central transition of a spin 3/2 nucleus:
$M(t)\!=\!M_0\left[1-2f\left(\frac{9}{10}e^{-(6t/T_1)^\beta} +\frac{1}{10}e^{-(t/T_1)^\beta}\right)\right]$
where $M_0$ is the equilibrium nuclear magnetization, $f$ is the inversion fraction, and $\beta$ is the stretched exponent. We find that $\beta\approx 0.7$ and is temperature-independent. Fig. \ref{fig:parameters}(c) displays $(T_1T)^{-1}$  as a function of temperature.  This quantity probes the dynamical spin susceptibility through the relationship:
\begin{equation}
	\label{eqn:dynamical_susceptibility}
	\left(\frac{1}{T_1T}\right)_{\alpha} = \gamma^2 k_B T \lim_{\omega \rightarrow 0} \sum\limits_{\mathbf{q},\beta\neq\alpha} \mathcal{F}_{\alpha\beta}(\mathbf{q}) \frac{\textrm{Im}\chi_{\alpha\beta}(\mathbf{q},\omega)}{\hslash \omega},
\end{equation}
where $\mathcal{F}_{\alpha\beta}(\mathbf{q})$ are form factors that depend on the hyperfine coupling tensor, $\chi_{\alpha\beta}(\mathbf{q},\omega)$ is  the dynamical magnetic susceptibility, and $\alpha,\beta = \left\{ x,y,z \right\}$.\cite{MoriyaT1formula}  The large peak in $(T_1T)^{-1}$ reflects the slowing down of critical fluctuations near $T_N$. As shown in the inset of Fig. \ref{fig:parameters}(c), $T_1T$ varies linearly, and a linear fit indicates this quantity vanishes at $T_N = 14.3$ K. Surprisingly, $(T_1T)^{-1}$ appears to be isotropic over the entire temperature range, despite the anisotropy observed in $K$ and the static susceptibility (Figs. \ref{fig:parameters}(b), \ref{fig:shift}).  Both anisotropy in the form factors and in the dynamical spin susceptibility itself can contribute to the anisotropy of $(T_1T)^{-1}$. However, the hyperfine couplings given in Table \ref{tab:hyperfinepars} vary by at most 20\% for the two field orientations, thus the critical spin fluctuations themselves must be largely isotropic in the paramagnetic state.  This result contrasts with NMR observations in $\alpha$-RuCl$_3$,\cite{BaekRuCl3PRL} but agree with magnetic x-ray scattering results in the paramagnetic state, where the zigzag correlations decrease isotropically with increasing the temperature.~\cite{HwanChun2015} The fact that the spin fluctuations in \niro\ show no difference between in-plane and out-of-plane magnetic fields suggests that, despite the strong spin-orbit coupling evident in the magnetic order, the same strong spin-orbit coupling conspires to produce a symmetry in the critical fluctuations that is indistinguishable from spherical symmetry, consistent with the isotropy of the Kitaev cubic axes.
\begin{figure}[h]
\begin{center}
\includegraphics[width=\linewidth]{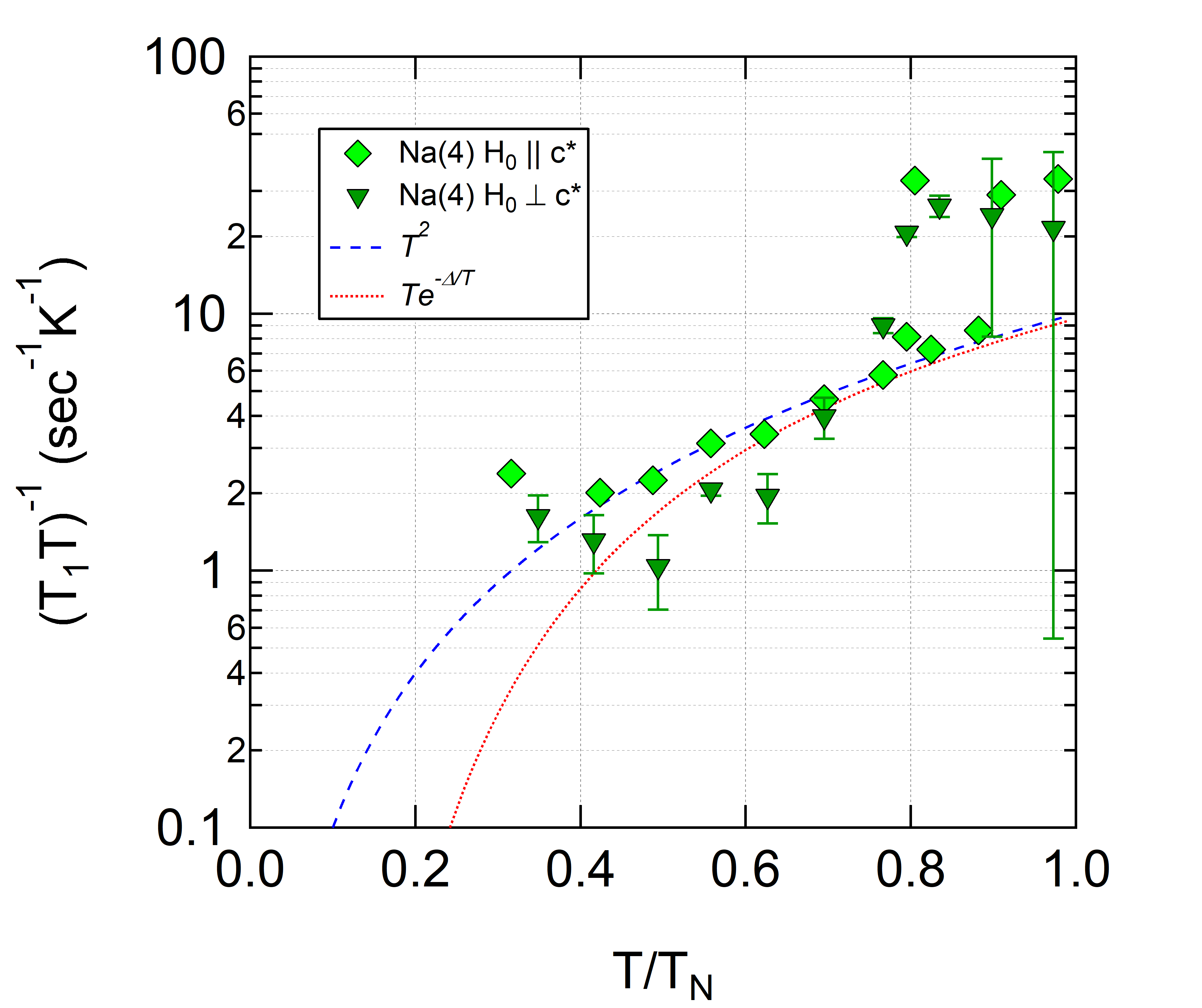}
\caption{$(T_1T)^{-1}$ versus $T$ for $\mathbf{H}_0 \parallel c^*$ ($\blacklozenge$) and $\mathbf{H}_0 \perp c^*$ ($\blacktriangledown$)  below  $T_N$. }
\label{t1t-AFM}
\end{center}
\end{figure}

Fig. \ref{t1t-AFM} displays $(T_1T)^{-1}$ versus temperature below $T_N$. It is well established that in a 3D conventional magnetic insulator the spin lattice relaxation rate exhibits either thermally activated behavior: $(T_1T)^{-1} \propto Te^{-\Delta/T}$, or power law relaxation: $(T_1T)^{-1} \propto T^{\alpha}$, where $\alpha = 2$ or $4$ for  two magnon or three magnon scattering. \cite{PhysRev.166.359,PhysRevB.90.184408} We fit the data to both expressions, but as seen in Fig. \ref{t1t-AFM} neither expression captures the behavior well. For an ideal Kitaev spin model, the NMR spin-lattice relaxation rate is expected to exhibit activated behavior: \slrr$\sim e^{-\Delta/k_BT}$, where $\Delta$ is the spin gap. \cite{MoessnerPRL,MoessnerPRB} For the more generic case of a gapless Kitaev quantum spin liquid on a honeycomb lattice, $T_1^{-1} \propto T^3$.~\cite{PhysRevLett.117.037209} However, neither of these cases adequately captures our observations either. Surprisingly, the data in Fig. \ref{t1t-AFM} approach a constant as $T\rightarrow0$.  This result is surprising because at $T=0$ all of the fluctuations should be frozen out.  $(T_1T)^{-1} = const$ suggests the presence of a finite density of states in a conductor, but \niro\ is a Mott insulator with a band gap of 300\,meV so relaxation by itinerant charges can be ruled out. Observations in similar materials reveal different trends. In polycrystalline samples of the 2D honeycomb material Li$_2$RhO$_3$, $T_1^{-1}\propto T^{2.2}$.~\cite{PhysRevB.96.094432}  In RuCl$_3$, \slrr\ exhibits qualitatively similar behavior at low fields, with a peak at $T_N$ and decreasing \slrr\ below; however at high fields \slrr\ is dramatically suppressed as the magnetism is suppressed and a spin gap emerges. \cite{BaekRuCl3PRL}

The presence of a small population of minority Ir spins located at the Na sites may  offer an explanation for the gapless excitations we observe at low temperatures in the antiferromagnetic phase, consistent with the breakdown of the $K-\chi$ relationship.  Because these minority spins may not order, they would continue to fluctuate and can contribute to the spin-lattice relaxation of the Na nuclei.  Substitutional disorder is not uncommon in these materials, and structural refinement studies have indicated that up to 35\% of the Na(2) sites can be occupied by Ir.\cite{CaoNa2IrO3zigzag}

An alternative possibility is that layer stacking faults may be present which could be correlated with each other in complicated ways. The net result is that some layers could magnetically interact with each other enough to give different regions of the crystal with somewhat different magnetic properties. Though they all undergo the same three dimensional $T_N$ transition, the resulting regions would have different magnetic sites across the crystal experiencing different internal fields. A related possibility is that stacking faults could produce inhomogeneous electric fields that can then change magnetic interaction energies; in quantum paramagnets such energy randomness can give rise to topological defects carrying spin 1/2 moments, with unusual low energy dynamics, and it is conceivable that some such magnetic defects can arise from energy randomness also in this strongly frustrated, albeit ordered, magnet.\cite{KimchiRandomQuantumMagnets,Kimchi2018}

In conclusion, we have found that the Na NMR spectrum exhibits a shift anomaly below $\sim 30$K, and a significant broadening below $T_N$ associated with a broad distribution of local internal fields in the antiferromagnetic state, despite independent observations of commensurate ordering.  We attribute these effects to Ir-Na substitutional disorder, which may give rise to a subset of Ir spins that exhibit different magnetic behavior than those in the honeycomb lattice. The surprising lack of anisotropy in \slrr\ in the paramagnetic phase suggests that isotropic spin fluctuations are driven by Kitaev interaction. Furthermore, \slrr\ data reveal gapless excitations deep in the ordered state, which may also arise from disorder. Although this type of substitutional disorder is particular to the A$_2$IrO$_3$ family, similar phenomena have been observed in RuCl$_3$ and it is important to account for the consequences of such a distribution into two subsystems to fully understand the nature of the low energy excitations within the quantum spin liquid phase of related materials.  Further studies at lower temperature may shed important light on the nature of these excitations.

\textit{Acknowledgment.} We acknowledge helpful discussions with T. Kissikov and B. Bush. This research is supported by the Deutsche Forschungsgemeinschaft (DFG) through SFB 1143 for the project C02. Work
at UC Davis was supported by the NSF under Grants Nos.\ DMR-1506961 and DMR-1807889. Work by JGA was supported by the Department of Energy, Office of Basic Energy Sciences, Early Career program under Contract No. DE-AC02-05CH11231.

\bibliography{Na2IrO3NMR}

\begin{thebibliography}{33}%
\makeatletter
\providecommand \@ifxundefined [1]{%
 \@ifx{#1\undefined}
}%
\providecommand \@ifnum [1]{%
 \ifnum #1\expandafter \@firstoftwo
 \else \expandafter \@secondoftwo
 \fi
}%
\providecommand \@ifx [1]{%
 \ifx #1\expandafter \@firstoftwo
 \else \expandafter \@secondoftwo
 \fi
}%
\providecommand \natexlab [1]{#1}%
\providecommand \enquote  [1]{``#1''}%
\providecommand \bibnamefont  [1]{#1}%
\providecommand \bibfnamefont [1]{#1}%
\providecommand \citenamefont [1]{#1}%
\providecommand \href@noop [0]{\@secondoftwo}%
\providecommand \href [0]{\begingroup \@sanitize@url \@href}%
\providecommand \@href[1]{\@@startlink{#1}\@@href}%
\providecommand \@@href[1]{\endgroup#1\@@endlink}%
\providecommand \@sanitize@url [0]{\catcode `\\12\catcode `\$12\catcode
  `\&12\catcode `\#12\catcode `\^12\catcode `\_12\catcode `\%12\relax}%
\providecommand \@@startlink[1]{}%
\providecommand \@@endlink[0]{}%
\providecommand \url  [0]{\begingroup\@sanitize@url \@url }%
\providecommand \@url [1]{\endgroup\@href {#1}{\urlprefix }}%
\providecommand \urlprefix  [0]{URL }%
\providecommand \Eprint [0]{\href }%
\providecommand \doibase [0]{http://dx.doi.org/}%
\providecommand \selectlanguage [0]{\@gobble}%
\providecommand \bibinfo  [0]{\@secondoftwo}%
\providecommand \bibfield  [0]{\@secondoftwo}%
\providecommand \translation [1]{[#1]}%
\providecommand \BibitemOpen [0]{}%
\providecommand \bibitemStop [0]{}%
\providecommand \bibitemNoStop [0]{.\EOS\space}%
\providecommand \EOS [0]{\spacefactor3000\relax}%
\providecommand \BibitemShut  [1]{\csname bibitem#1\endcsname}%
\let\auto@bib@innerbib\@empty
\bibitem [{\citenamefont {Trebst}(2017)}]{KitaevMaterialsReview}%
  \BibitemOpen
  \bibfield  {author} {\bibinfo {author} {\bibfnamefont {Simon}\ \bibnamefont
  {Trebst}},\ }\bibfield  {title} {\enquote {\bibinfo {title} {Kitaev
  materials},}\ }\href@noop {} {\  (\bibinfo {year} {2017})},\ \Eprint
  {http://arxiv.org/abs/1701.07056v1} {1701.07056v1} \BibitemShut {NoStop}%
\bibitem [{\citenamefont {Gegenwart}\ and\ \citenamefont
  {Trebst}(2015)}]{Gegenwart2015}%
  \BibitemOpen
  \bibfield  {author} {\bibinfo {author} {\bibfnamefont {Philipp}\ \bibnamefont
  {Gegenwart}}\ and\ \bibinfo {author} {\bibfnamefont {Simon}\ \bibnamefont
  {Trebst}},\ }\bibfield  {title} {\enquote {\bibinfo {title} {Kitaev
  matter},}\ }\href {\doibase 10.1038/nphys3346} {\bibfield  {journal}
  {\bibinfo  {journal} {Nature Physics}\ }\textbf {\bibinfo {volume} {11}},\
  \bibinfo {pages} {444--445} (\bibinfo {year} {2015})}\BibitemShut {NoStop}%
\bibitem [{\citenamefont {Hermanns}\ \emph {et~al.}(2018)\citenamefont
  {Hermanns}, \citenamefont {Kimchi},\ and\ \citenamefont
  {Knolle}}]{KimchiKitaevReview}%
  \BibitemOpen
  \bibfield  {author} {\bibinfo {author} {\bibfnamefont {M.}~\bibnamefont
  {Hermanns}}, \bibinfo {author} {\bibfnamefont {I.}~\bibnamefont {Kimchi}}, \
  and\ \bibinfo {author} {\bibfnamefont {J.}~\bibnamefont {Knolle}},\
  }\bibfield  {title} {\enquote {\bibinfo {title} {Physics of the kitaev model:
  Fractionalization, dynamic correlations, and material connections},}\ }\href
  {\doibase 10.1146/annurev-conmatphys-033117-053934} {\bibfield  {journal}
  {\bibinfo  {journal} {Annual Review of Condensed Matter Physics}\ }\textbf
  {\bibinfo {volume} {9}},\ \bibinfo {pages} {17--33} (\bibinfo {year}
  {2018})}\BibitemShut {NoStop}%
\bibitem [{\citenamefont {Jackeli}\ and\ \citenamefont
  {Khaliullin}(2009)}]{JackeliKhaliullin}%
  \BibitemOpen
  \bibfield  {author} {\bibinfo {author} {\bibfnamefont {G.}~\bibnamefont
  {Jackeli}}\ and\ \bibinfo {author} {\bibfnamefont {G.}~\bibnamefont
  {Khaliullin}},\ }\bibfield  {title} {\enquote {\bibinfo {title} {Mott
  insulators in the strong spin-orbit coupling limit: From {H}eisenberg to a
  quantum compass and {K}itaev models},}\ }\href {\doibase
  10.1103/PhysRevLett.102.017205} {\bibfield  {journal} {\bibinfo  {journal}
  {Phys. Rev. Lett.}\ }\textbf {\bibinfo {volume} {102}},\ \bibinfo {pages}
  {017205} (\bibinfo {year} {2009})}\BibitemShut {NoStop}%
\bibitem [{\citenamefont {Takagi}\ \emph {et~al.}(2019)\citenamefont {Takagi},
  \citenamefont {Takayama}, \citenamefont {Jackeli}, \citenamefont
  {Khaliullin},\ and\ \citenamefont {Nagler}}]{Takagi2019}%
  \BibitemOpen
  \bibfield  {author} {\bibinfo {author} {\bibfnamefont {Hidenori}\
  \bibnamefont {Takagi}}, \bibinfo {author} {\bibfnamefont {Tomohiro}\
  \bibnamefont {Takayama}}, \bibinfo {author} {\bibfnamefont {George}\
  \bibnamefont {Jackeli}}, \bibinfo {author} {\bibfnamefont {Giniyat}\
  \bibnamefont {Khaliullin}}, \ and\ \bibinfo {author} {\bibfnamefont
  {Stephen~E.}\ \bibnamefont {Nagler}},\ }\bibfield  {title} {\enquote
  {\bibinfo {title} {Concept and realization of {K}itaev quantum spin
  liquids},}\ }\href {\doibase 10.1038/s42254-019-0038-2} {\bibfield  {journal}
  {\bibinfo  {journal} {Nature Reviews Physics}\ }\textbf {\bibinfo {volume}
  {1}},\ \bibinfo {pages} {264--280} (\bibinfo {year} {2019})}\BibitemShut
  {NoStop}%
\bibitem [{\citenamefont {Comin}\ \emph {et~al.}(2012)\citenamefont {Comin},
  \citenamefont {Levy}, \citenamefont {Ludbrook}, \citenamefont {Zhu},
  \citenamefont {Veenstra}, \citenamefont {Rosen}, \citenamefont {Singh},
  \citenamefont {Gegenwart}, \citenamefont {Stricker}, \citenamefont {Hancock},
  \citenamefont {van~der Marel}, \citenamefont {Elfimov},\ and\ \citenamefont
  {Damascelli}}]{CominNa2IrO3}%
  \BibitemOpen
  \bibfield  {author} {\bibinfo {author} {\bibfnamefont {R.}~\bibnamefont
  {Comin}}, \bibinfo {author} {\bibfnamefont {G.}~\bibnamefont {Levy}},
  \bibinfo {author} {\bibfnamefont {B.}~\bibnamefont {Ludbrook}}, \bibinfo
  {author} {\bibfnamefont {Z.-H.}\ \bibnamefont {Zhu}}, \bibinfo {author}
  {\bibfnamefont {C.~N.}\ \bibnamefont {Veenstra}}, \bibinfo {author}
  {\bibfnamefont {J.~A.}\ \bibnamefont {Rosen}}, \bibinfo {author}
  {\bibfnamefont {Yogesh}\ \bibnamefont {Singh}}, \bibinfo {author}
  {\bibfnamefont {P.}~\bibnamefont {Gegenwart}}, \bibinfo {author}
  {\bibfnamefont {D.}~\bibnamefont {Stricker}}, \bibinfo {author}
  {\bibfnamefont {J.~N.}\ \bibnamefont {Hancock}}, \bibinfo {author}
  {\bibfnamefont {D.}~\bibnamefont {van~der Marel}}, \bibinfo {author}
  {\bibfnamefont {I.~S.}\ \bibnamefont {Elfimov}}, \ and\ \bibinfo {author}
  {\bibfnamefont {A.}~\bibnamefont {Damascelli}},\ }\bibfield  {title}
  {\enquote {\bibinfo {title} {{Na$_2$IrO$_3$} as a novel relativistic {M}ott
  insulator with a 340-mev gap},}\ }\href {\doibase
  10.1103/PhysRevLett.109.266406} {\bibfield  {journal} {\bibinfo  {journal}
  {Phys. Rev. Lett.}\ }\textbf {\bibinfo {volume} {109}},\ \bibinfo {pages}
  {266406} (\bibinfo {year} {2012})}\BibitemShut {NoStop}%
\bibitem [{\citenamefont {Chaloupka}\ \emph {et~al.}(2013)\citenamefont
  {Chaloupka}, \citenamefont {Jackeli},\ and\ \citenamefont
  {Khaliullin}}]{ZigZagNaIrO3}%
  \BibitemOpen
  \bibfield  {author} {\bibinfo {author} {\bibfnamefont {Ji\ifmmode
  \check{r}\else~\v{r}\fi{}\'{\i}}\ \bibnamefont {Chaloupka}}, \bibinfo
  {author} {\bibfnamefont {George}\ \bibnamefont {Jackeli}}, \ and\ \bibinfo
  {author} {\bibfnamefont {Giniyat}\ \bibnamefont {Khaliullin}},\ }\bibfield
  {title} {\enquote {\bibinfo {title} {Zigzag magnetic order in the iridium
  oxide {Na$_2$IrO$_3$}},}\ }\href {\doibase 10.1103/PhysRevLett.110.097204}
  {\bibfield  {journal} {\bibinfo  {journal} {Phys. Rev. Lett.}\ }\textbf
  {\bibinfo {volume} {110}},\ \bibinfo {pages} {097204} (\bibinfo {year}
  {2013})}\BibitemShut {NoStop}%
\bibitem [{\citenamefont {Liu}\ \emph {et~al.}(2011)\citenamefont {Liu},
  \citenamefont {Berlijn}, \citenamefont {Yin}, \citenamefont {Ku},
  \citenamefont {Tsvelik}, \citenamefont {Kim}, \citenamefont {Gretarsson},
  \citenamefont {Singh}, \citenamefont {Gegenwart},\ and\ \citenamefont
  {Hill}}]{GegenwartNa2IrO3PRB2011}%
  \BibitemOpen
  \bibfield  {author} {\bibinfo {author} {\bibfnamefont {X.}~\bibnamefont
  {Liu}}, \bibinfo {author} {\bibfnamefont {T.}~\bibnamefont {Berlijn}},
  \bibinfo {author} {\bibfnamefont {W.-G.}\ \bibnamefont {Yin}}, \bibinfo
  {author} {\bibfnamefont {W.}~\bibnamefont {Ku}}, \bibinfo {author}
  {\bibfnamefont {A.}~\bibnamefont {Tsvelik}}, \bibinfo {author} {\bibfnamefont
  {Young-June}\ \bibnamefont {Kim}}, \bibinfo {author} {\bibfnamefont
  {H.}~\bibnamefont {Gretarsson}}, \bibinfo {author} {\bibfnamefont {Yogesh}\
  \bibnamefont {Singh}}, \bibinfo {author} {\bibfnamefont {P.}~\bibnamefont
  {Gegenwart}}, \ and\ \bibinfo {author} {\bibfnamefont {J.~P.}\ \bibnamefont
  {Hill}},\ }\bibfield  {title} {\enquote {\bibinfo {title} {Long-range
  magnetic ordering in {Na$_2$IrO$_3$}},}\ }\href {\doibase
  10.1103/PhysRevB.83.220403} {\bibfield  {journal} {\bibinfo  {journal} {Phys.
  Rev. B}\ }\textbf {\bibinfo {volume} {83}},\ \bibinfo {pages} {220403}
  (\bibinfo {year} {2011})}\BibitemShut {NoStop}%
\bibitem [{\citenamefont {Ye}\ \emph {et~al.}(2012)\citenamefont {Ye},
  \citenamefont {Chi}, \citenamefont {Cao}, \citenamefont {Chakoumakos},
  \citenamefont {Fernandez-Baca}, \citenamefont {Custelcean}, \citenamefont
  {Qi}, \citenamefont {Korneta},\ and\ \citenamefont {Cao}}]{CaoNa2IrO3zigzag}%
  \BibitemOpen
  \bibfield  {author} {\bibinfo {author} {\bibfnamefont {Feng}\ \bibnamefont
  {Ye}}, \bibinfo {author} {\bibfnamefont {Songxue}\ \bibnamefont {Chi}},
  \bibinfo {author} {\bibfnamefont {Huibo}\ \bibnamefont {Cao}}, \bibinfo
  {author} {\bibfnamefont {Bryan~C.}\ \bibnamefont {Chakoumakos}}, \bibinfo
  {author} {\bibfnamefont {Jaime~A.}\ \bibnamefont {Fernandez-Baca}}, \bibinfo
  {author} {\bibfnamefont {Radu}\ \bibnamefont {Custelcean}}, \bibinfo {author}
  {\bibfnamefont {T.~F.}\ \bibnamefont {Qi}}, \bibinfo {author} {\bibfnamefont
  {O.~B.}\ \bibnamefont {Korneta}}, \ and\ \bibinfo {author} {\bibfnamefont
  {G.}~\bibnamefont {Cao}},\ }\bibfield  {title} {\enquote {\bibinfo {title}
  {Direct evidence of a zigzag spin-chain structure in the honeycomb lattice: A
  neutron and x-ray diffraction investigation of single-crystal
  {Na$_2$IrO$_3$}},}\ }\href {\doibase 10.1103/PhysRevB.85.180403} {\bibfield
  {journal} {\bibinfo  {journal} {Phys. Rev. B}\ }\textbf {\bibinfo {volume}
  {85}},\ \bibinfo {pages} {180403} (\bibinfo {year} {2012})}\BibitemShut
  {NoStop}%
\bibitem [{\citenamefont {Singh}\ \emph {et~al.}(2012)\citenamefont {Singh},
  \citenamefont {Manni}, \citenamefont {Reuther}, \citenamefont {Berlijn},
  \citenamefont {Thomale}, \citenamefont {Ku}, \citenamefont {Trebst},\ and\
  \citenamefont {Gegenwart}}]{KitaevA2IrO3PRL}%
  \BibitemOpen
  \bibfield  {author} {\bibinfo {author} {\bibfnamefont {Yogesh}\ \bibnamefont
  {Singh}}, \bibinfo {author} {\bibfnamefont {S.}~\bibnamefont {Manni}},
  \bibinfo {author} {\bibfnamefont {J.}~\bibnamefont {Reuther}}, \bibinfo
  {author} {\bibfnamefont {T.}~\bibnamefont {Berlijn}}, \bibinfo {author}
  {\bibfnamefont {R.}~\bibnamefont {Thomale}}, \bibinfo {author} {\bibfnamefont
  {W.}~\bibnamefont {Ku}}, \bibinfo {author} {\bibfnamefont {S.}~\bibnamefont
  {Trebst}}, \ and\ \bibinfo {author} {\bibfnamefont {P.}~\bibnamefont
  {Gegenwart}},\ }\bibfield  {title} {\enquote {\bibinfo {title} {Relevance of
  the {Heisenberg-Kitaev} model for the honeycomb lattice iridates
  {A$_2$IrO$_3$}},}\ }\href {\doibase 10.1103/PhysRevLett.108.127203}
  {\bibfield  {journal} {\bibinfo  {journal} {Phys. Rev. Lett.}\ }\textbf
  {\bibinfo {volume} {108}},\ \bibinfo {pages} {127203} (\bibinfo {year}
  {2012})}\BibitemShut {NoStop}%
\bibitem [{\citenamefont {Singh}\ and\ \citenamefont
  {Gegenwart}(2010)}]{SinghGegenwart}%
  \BibitemOpen
  \bibfield  {author} {\bibinfo {author} {\bibfnamefont {Yogesh}\ \bibnamefont
  {Singh}}\ and\ \bibinfo {author} {\bibfnamefont {P.}~\bibnamefont
  {Gegenwart}},\ }\bibfield  {title} {\enquote {\bibinfo {title}
  {Antiferromagnetic mott insulating state in single crystals of the honeycomb
  lattice material {Na$_2$IrO$_3$}},}\ }\href {\doibase
  10.1103/PhysRevB.82.064412} {\bibfield  {journal} {\bibinfo  {journal} {Phys.
  Rev. B}\ }\textbf {\bibinfo {volume} {82}},\ \bibinfo {pages} {064412}
  (\bibinfo {year} {2010})}\BibitemShut {NoStop}%
\bibitem [{\citenamefont {Hwan~Chun}\ \emph {et~al.}(2015)\citenamefont
  {Hwan~Chun}, \citenamefont {Kim}, \citenamefont {Kim}, \citenamefont {Zheng},
  \citenamefont {Stoumpos}, \citenamefont {Malliakas}, \citenamefont
  {Mitchell}, \citenamefont {Mehlawat}, \citenamefont {Singh}, \citenamefont
  {Choi}, \citenamefont {Gog}, \citenamefont {Al-Zein}, \citenamefont {Sala},
  \citenamefont {Krisch}, \citenamefont {Chaloupka}, \citenamefont {Jackeli},
  \citenamefont {Khaliullin},\ and\ \citenamefont {Kim}}]{HwanChun2015}%
  \BibitemOpen
  \bibfield  {author} {\bibinfo {author} {\bibfnamefont {Sae}\ \bibnamefont
  {Hwan~Chun}}, \bibinfo {author} {\bibfnamefont {Jong-Woo}\ \bibnamefont
  {Kim}}, \bibinfo {author} {\bibfnamefont {Jungho}\ \bibnamefont {Kim}},
  \bibinfo {author} {\bibfnamefont {H.}~\bibnamefont {Zheng}}, \bibinfo
  {author} {\bibfnamefont {Constantinos~C.}\ \bibnamefont {Stoumpos}}, \bibinfo
  {author} {\bibfnamefont {C.~D.}\ \bibnamefont {Malliakas}}, \bibinfo {author}
  {\bibfnamefont {J.~F.}\ \bibnamefont {Mitchell}}, \bibinfo {author}
  {\bibfnamefont {Kavita}\ \bibnamefont {Mehlawat}}, \bibinfo {author}
  {\bibfnamefont {Yogesh}\ \bibnamefont {Singh}}, \bibinfo {author}
  {\bibfnamefont {Y.}~\bibnamefont {Choi}}, \bibinfo {author} {\bibfnamefont
  {T.}~\bibnamefont {Gog}}, \bibinfo {author} {\bibfnamefont {A.}~\bibnamefont
  {Al-Zein}}, \bibinfo {author} {\bibfnamefont {M.~Moretti}\ \bibnamefont
  {Sala}}, \bibinfo {author} {\bibfnamefont {M.}~\bibnamefont {Krisch}},
  \bibinfo {author} {\bibfnamefont {J.}~\bibnamefont {Chaloupka}}, \bibinfo
  {author} {\bibfnamefont {G.}~\bibnamefont {Jackeli}}, \bibinfo {author}
  {\bibfnamefont {G.}~\bibnamefont {Khaliullin}}, \ and\ \bibinfo {author}
  {\bibfnamefont {B.~J.}\ \bibnamefont {Kim}},\ }\bibfield  {title} {\enquote
  {\bibinfo {title} {Direct evidence for dominant bond-directional interactions
  in a honeycomb lattice iridate {Na$_2$IrO$_3$}},}\ }\href
  {http://dx.doi.org/10.1038/nphys3322} {\bibfield  {journal} {\bibinfo
  {journal} {Nat Phys}\ }\textbf {\bibinfo {volume} {11}},\ \bibinfo {pages}
  {462--466} (\bibinfo {year} {2015})}\BibitemShut {NoStop}%
\bibitem [{\citenamefont {Baek}\ \emph {et~al.}(2017)\citenamefont {Baek},
  \citenamefont {Do}, \citenamefont {Choi}, \citenamefont {Kwon}, \citenamefont
  {Wolter}, \citenamefont {Nishimoto}, \citenamefont {van~den Brink},\ and\
  \citenamefont {B\"uchner}}]{BaekRuCl3PRL}%
  \BibitemOpen
  \bibfield  {author} {\bibinfo {author} {\bibfnamefont {S.-H.}\ \bibnamefont
  {Baek}}, \bibinfo {author} {\bibfnamefont {S.-H.}\ \bibnamefont {Do}},
  \bibinfo {author} {\bibfnamefont {K.-Y.}\ \bibnamefont {Choi}}, \bibinfo
  {author} {\bibfnamefont {Y.~S.}\ \bibnamefont {Kwon}}, \bibinfo {author}
  {\bibfnamefont {A.~U.~B.}\ \bibnamefont {Wolter}}, \bibinfo {author}
  {\bibfnamefont {S.}~\bibnamefont {Nishimoto}}, \bibinfo {author}
  {\bibfnamefont {Jeroen}\ \bibnamefont {van~den Brink}}, \ and\ \bibinfo
  {author} {\bibfnamefont {B.}~\bibnamefont {B\"uchner}},\ }\bibfield  {title}
  {\enquote {\bibinfo {title} {Evidence for a field-induced quantum spin liquid
  in {$\ensuremath{\alpha}$-${\mathrm{RuCl}}_{3}$}},}\ }\href {\doibase
  10.1103/PhysRevLett.119.037201} {\bibfield  {journal} {\bibinfo  {journal}
  {Phys. Rev. Lett.}\ }\textbf {\bibinfo {volume} {119}},\ \bibinfo {pages}
  {037201} (\bibinfo {year} {2017})}\BibitemShut {NoStop}%
\bibitem [{\citenamefont {Jan{\v{s}}a}\ \emph {et~al.}(2018)\citenamefont
  {Jan{\v{s}}a}, \citenamefont {Zorko}, \citenamefont {Gomil{\v{s}}ek},
  \citenamefont {Pregelj}, \citenamefont {Kr\"{a}mer}, \citenamefont {Biner},
  \citenamefont {Biffin}, \citenamefont {R\"{u}egg},\ and\ \citenamefont
  {Klanj{\v{s}}ek}}]{Jansa2018}%
  \BibitemOpen
  \bibfield  {author} {\bibinfo {author} {\bibfnamefont {Nejc}\ \bibnamefont
  {Jan{\v{s}}a}}, \bibinfo {author} {\bibfnamefont {Andrej}\ \bibnamefont
  {Zorko}}, \bibinfo {author} {\bibfnamefont {Matja{\v{z}}}\ \bibnamefont
  {Gomil{\v{s}}ek}}, \bibinfo {author} {\bibfnamefont {Matej}\ \bibnamefont
  {Pregelj}}, \bibinfo {author} {\bibfnamefont {Karl~W.}\ \bibnamefont
  {Kr\"{a}mer}}, \bibinfo {author} {\bibfnamefont {Daniel}\ \bibnamefont
  {Biner}}, \bibinfo {author} {\bibfnamefont {Alun}\ \bibnamefont {Biffin}},
  \bibinfo {author} {\bibfnamefont {Christian}\ \bibnamefont {R\"{u}egg}}, \
  and\ \bibinfo {author} {\bibfnamefont {Martin}\ \bibnamefont
  {Klanj{\v{s}}ek}},\ }\bibfield  {title} {\enquote {\bibinfo {title}
  {Observation of two types of fractional excitation in the {K}itaev honeycomb
  magnet},}\ }\href {\doibase 10.1038/s41567-018-0129-5} {\bibfield  {journal}
  {\bibinfo  {journal} {Nature Physics}\ }\textbf {\bibinfo {volume} {14}},\
  \bibinfo {pages} {786--790} (\bibinfo {year} {2018})}\BibitemShut {NoStop}%
\bibitem [{\citenamefont {Zheng}\ \emph {et~al.}(2017)\citenamefont {Zheng},
  \citenamefont {Ran}, \citenamefont {Li}, \citenamefont {Wang}, \citenamefont
  {Wang}, \citenamefont {Liu}, \citenamefont {Liu}, \citenamefont {Normand},
  \citenamefont {Wen},\ and\ \citenamefont {Yu}}]{ZhengRuCl3PRL}%
  \BibitemOpen
  \bibfield  {author} {\bibinfo {author} {\bibfnamefont {Jiacheng}\
  \bibnamefont {Zheng}}, \bibinfo {author} {\bibfnamefont {Kejing}\
  \bibnamefont {Ran}}, \bibinfo {author} {\bibfnamefont {Tianrun}\ \bibnamefont
  {Li}}, \bibinfo {author} {\bibfnamefont {Jinghui}\ \bibnamefont {Wang}},
  \bibinfo {author} {\bibfnamefont {Pengshuai}\ \bibnamefont {Wang}}, \bibinfo
  {author} {\bibfnamefont {Bin}\ \bibnamefont {Liu}}, \bibinfo {author}
  {\bibfnamefont {Zheng-Xin}\ \bibnamefont {Liu}}, \bibinfo {author}
  {\bibfnamefont {B.}~\bibnamefont {Normand}}, \bibinfo {author} {\bibfnamefont
  {Jinsheng}\ \bibnamefont {Wen}}, \ and\ \bibinfo {author} {\bibfnamefont
  {Weiqiang}\ \bibnamefont {Yu}},\ }\bibfield  {title} {\enquote {\bibinfo
  {title} {Gapless spin excitations in the field-induced quantum spin liquid
  phase of {$\ensuremath{\alpha}\text{\ensuremath{-}}{\mathrm{RuCl}}_{3}$}},}\
  }\href {\doibase 10.1103/PhysRevLett.119.227208} {\bibfield  {journal}
  {\bibinfo  {journal} {Phys. Rev. Lett.}\ }\textbf {\bibinfo {volume} {119}},\
  \bibinfo {pages} {227208} (\bibinfo {year} {2017})}\BibitemShut {NoStop}%
\bibitem [{\citenamefont {Kitagawa}\ \emph {et~al.}(2018)\citenamefont
  {Kitagawa}, \citenamefont {Takayama}, \citenamefont {Matsumoto},
  \citenamefont {Kato}, \citenamefont {Takano}, \citenamefont {Kishimoto},
  \citenamefont {Bette}, \citenamefont {Dinnebier}, \citenamefont {Jackeli},\
  and\ \citenamefont {Takagi}}]{Kitagawa2018}%
  \BibitemOpen
  \bibfield  {author} {\bibinfo {author} {\bibfnamefont {K.}~\bibnamefont
  {Kitagawa}}, \bibinfo {author} {\bibfnamefont {T.}~\bibnamefont {Takayama}},
  \bibinfo {author} {\bibfnamefont {Y.}~\bibnamefont {Matsumoto}}, \bibinfo
  {author} {\bibfnamefont {A.}~\bibnamefont {Kato}}, \bibinfo {author}
  {\bibfnamefont {R.}~\bibnamefont {Takano}}, \bibinfo {author} {\bibfnamefont
  {Y.}~\bibnamefont {Kishimoto}}, \bibinfo {author} {\bibfnamefont
  {S.}~\bibnamefont {Bette}}, \bibinfo {author} {\bibfnamefont
  {R.}~\bibnamefont {Dinnebier}}, \bibinfo {author} {\bibfnamefont
  {G.}~\bibnamefont {Jackeli}}, \ and\ \bibinfo {author} {\bibfnamefont
  {H.}~\bibnamefont {Takagi}},\ }\bibfield  {title} {\enquote {\bibinfo {title}
  {A spin{\textendash}orbital-entangled quantum liquid on a honeycomb
  lattice},}\ }\href {\doibase 10.1038/nature25482} {\bibfield  {journal}
  {\bibinfo  {journal} {Nature}\ }\textbf {\bibinfo {volume} {554}},\ \bibinfo
  {pages} {341--345} (\bibinfo {year} {2018})}\BibitemShut {NoStop}%
\bibitem [{\citenamefont {Majumder}\ \emph {et~al.}(2019)\citenamefont
  {Majumder}, \citenamefont {Freund}, \citenamefont {Dey}, \citenamefont
  {Prinz-Zwick}, \citenamefont {B\"uttgen}, \citenamefont {Skourski},
  \citenamefont {Jesche}, \citenamefont {Tsirlin},\ and\ \citenamefont
  {Gegenwart}}]{Li2IrO3NMR2019}%
  \BibitemOpen
  \bibfield  {author} {\bibinfo {author} {\bibfnamefont {M.}~\bibnamefont
  {Majumder}}, \bibinfo {author} {\bibfnamefont {F.}~\bibnamefont {Freund}},
  \bibinfo {author} {\bibfnamefont {T.}~\bibnamefont {Dey}}, \bibinfo {author}
  {\bibfnamefont {M.}~\bibnamefont {Prinz-Zwick}}, \bibinfo {author}
  {\bibfnamefont {N.}~\bibnamefont {B\"uttgen}}, \bibinfo {author}
  {\bibfnamefont {Y.}~\bibnamefont {Skourski}}, \bibinfo {author}
  {\bibfnamefont {A.}~\bibnamefont {Jesche}}, \bibinfo {author} {\bibfnamefont
  {A.~A.}\ \bibnamefont {Tsirlin}}, \ and\ \bibinfo {author} {\bibfnamefont
  {P.}~\bibnamefont {Gegenwart}},\ }\bibfield  {title} {\enquote {\bibinfo
  {title} {Anisotropic temperature-field phase diagram of single crystalline
  {$\ensuremath{\beta}\ensuremath{-}{\mathrm{Li}}_{2}{\mathrm{IrO}}_{3}$}:
  Magnetization, specific heat, and {$^{7}\mathrm{Li}$ NMR} study},}\ }\href
  {\doibase 10.1103/PhysRevMaterials.3.074408} {\bibfield  {journal} {\bibinfo
  {journal} {Phys. Rev. Materials}\ }\textbf {\bibinfo {volume} {3}},\ \bibinfo
  {pages} {074408} (\bibinfo {year} {2019})}\BibitemShut {NoStop}%
\bibitem [{\citenamefont {Simutis}\ \emph {et~al.}(2018)\citenamefont
  {Simutis}, \citenamefont {Barbero}, \citenamefont {Rolfs}, \citenamefont
  {Leroy-Calatayud}, \citenamefont {Mehlawat}, \citenamefont {Khasanov},
  \citenamefont {Luetkens}, \citenamefont {Pomjakushina}, \citenamefont
  {Singh}, \citenamefont {Ott}, \citenamefont {Mesot}, \citenamefont {Amato},\
  and\ \citenamefont {Shiroka}}]{Simutis2018}%
  \BibitemOpen
  \bibfield  {author} {\bibinfo {author} {\bibfnamefont {G.}~\bibnamefont
  {Simutis}}, \bibinfo {author} {\bibfnamefont {N.}~\bibnamefont {Barbero}},
  \bibinfo {author} {\bibfnamefont {K.}~\bibnamefont {Rolfs}}, \bibinfo
  {author} {\bibfnamefont {P.}~\bibnamefont {Leroy-Calatayud}}, \bibinfo
  {author} {\bibfnamefont {K.}~\bibnamefont {Mehlawat}}, \bibinfo {author}
  {\bibfnamefont {R.}~\bibnamefont {Khasanov}}, \bibinfo {author}
  {\bibfnamefont {H.}~\bibnamefont {Luetkens}}, \bibinfo {author}
  {\bibfnamefont {E.}~\bibnamefont {Pomjakushina}}, \bibinfo {author}
  {\bibfnamefont {Y.}~\bibnamefont {Singh}}, \bibinfo {author} {\bibfnamefont
  {H.-R.}\ \bibnamefont {Ott}}, \bibinfo {author} {\bibfnamefont
  {J.}~\bibnamefont {Mesot}}, \bibinfo {author} {\bibfnamefont
  {A.}~\bibnamefont {Amato}}, \ and\ \bibinfo {author} {\bibfnamefont
  {T.}~\bibnamefont {Shiroka}},\ }\bibfield  {title} {\enquote {\bibinfo
  {title} {Chemical and hydrostatic-pressure effects on the kitaev honeycomb
  material {Na$_2$IrO$_3$}},}\ }\href {\doibase 10.1103/physrevb.98.104421}
  {\bibfield  {journal} {\bibinfo  {journal} {Phys. Rev. B}\ }\textbf {\bibinfo
  {volume} {98}},\ \bibinfo {pages} {104421} (\bibinfo {year}
  {2018})}\BibitemShut {NoStop}%
\bibitem [{\citenamefont {Slichter}(1992)}]{CPSbook}%
  \BibitemOpen
  \bibfield  {author} {\bibinfo {author} {\bibfnamefont {C.~P.}\ \bibnamefont
  {Slichter}},\ }\href@noop {} {\emph {\bibinfo {title} {Principles of Nuclear
  Magnetic Resonance}}},\ \bibinfo {edition} {3rd}\ ed.\ (\bibinfo  {publisher}
  {Springer-Verlag},\ \bibinfo {year} {1992})\BibitemShut {NoStop}%
\bibitem [{\citenamefont {Choi}\ \emph {et~al.}(2012)\citenamefont {Choi},
  \citenamefont {Coldea}, \citenamefont {Kolmogorov}, \citenamefont
  {Lancaster}, \citenamefont {Mazin}, \citenamefont {Blundell}, \citenamefont
  {Radaelli}, \citenamefont {Singh}, \citenamefont {Gegenwart}, \citenamefont
  {Choi}, \citenamefont {Cheong}, \citenamefont {Baker}, \citenamefont
  {Stock},\ and\ \citenamefont {Taylor}}]{Na2IrO3muSR}%
  \BibitemOpen
  \bibfield  {author} {\bibinfo {author} {\bibfnamefont {S.~K.}\ \bibnamefont
  {Choi}}, \bibinfo {author} {\bibfnamefont {R.}~\bibnamefont {Coldea}},
  \bibinfo {author} {\bibfnamefont {A.~N.}\ \bibnamefont {Kolmogorov}},
  \bibinfo {author} {\bibfnamefont {T.}~\bibnamefont {Lancaster}}, \bibinfo
  {author} {\bibfnamefont {I.~I.}\ \bibnamefont {Mazin}}, \bibinfo {author}
  {\bibfnamefont {S.~J.}\ \bibnamefont {Blundell}}, \bibinfo {author}
  {\bibfnamefont {P.~G.}\ \bibnamefont {Radaelli}}, \bibinfo {author}
  {\bibfnamefont {Yogesh}\ \bibnamefont {Singh}}, \bibinfo {author}
  {\bibfnamefont {P.}~\bibnamefont {Gegenwart}}, \bibinfo {author}
  {\bibfnamefont {K.~R.}\ \bibnamefont {Choi}}, \bibinfo {author}
  {\bibfnamefont {S.-W.}\ \bibnamefont {Cheong}}, \bibinfo {author}
  {\bibfnamefont {P.~J.}\ \bibnamefont {Baker}}, \bibinfo {author}
  {\bibfnamefont {C.}~\bibnamefont {Stock}}, \ and\ \bibinfo {author}
  {\bibfnamefont {J.}~\bibnamefont {Taylor}},\ }\bibfield  {title} {\enquote
  {\bibinfo {title} {Spin waves and revised crystal structure of honeycomb
  iridate {Na$_2$IrO$_3$}},}\ }\href {\doibase 10.1103/PhysRevLett.108.127204}
  {\bibfield  {journal} {\bibinfo  {journal} {Phys. Rev. Lett.}\ }\textbf
  {\bibinfo {volume} {108}},\ \bibinfo {pages} {127204} (\bibinfo {year}
  {2012})}\BibitemShut {NoStop}%
\bibitem [{\citenamefont {Dioguardi}\ \emph {et~al.}(2010)\citenamefont
  {Dioguardi}, \citenamefont {apRoberts Warren}, \citenamefont {Shockley},
  \citenamefont {Bud'ko}, \citenamefont {Ni}, \citenamefont {Canfield},\ and\
  \citenamefont {Curro}}]{Dioguardi2010}%
  \BibitemOpen
  \bibfield  {author} {\bibinfo {author} {\bibfnamefont {A.~P.}\ \bibnamefont
  {Dioguardi}}, \bibinfo {author} {\bibfnamefont {N.}~\bibnamefont {apRoberts
  Warren}}, \bibinfo {author} {\bibfnamefont {A.~C.}\ \bibnamefont {Shockley}},
  \bibinfo {author} {\bibfnamefont {S.~L.}\ \bibnamefont {Bud'ko}}, \bibinfo
  {author} {\bibfnamefont {N.}~\bibnamefont {Ni}}, \bibinfo {author}
  {\bibfnamefont {P.~C.}\ \bibnamefont {Canfield}}, \ and\ \bibinfo {author}
  {\bibfnamefont {N.~J.}\ \bibnamefont {Curro}},\ }\bibfield  {title} {\enquote
  {\bibinfo {title} {Local magnetic inhomogeneities in
  {Ba(Fe$_{1-x}$Ni$_x$)$_2$As$_2$} as seen via {As-75} {NMR}},}\ }\href
  {\doibase 10.1103/PhysRevB.82.140411} {\bibfield  {journal} {\bibinfo
  {journal} {Phys. Rev. B}\ }\textbf {\bibinfo {volume} {82}},\ \bibinfo
  {pages} {140411(R)} (\bibinfo {year} {2010})}\BibitemShut {NoStop}%
\bibitem [{\citenamefont {Curro}(2014)}]{Curro2012}%
  \BibitemOpen
  \bibfield  {author} {\bibinfo {author} {\bibfnamefont {Nicholas~J}\
  \bibnamefont {Curro}},\ }\enquote {\bibinfo {title} {Nuclear magnetic
  resonance as a probe of strongly correlated electron systems},}\ \ (\bibinfo
  {publisher} {Springer},\ \bibinfo {year} {2014})\ pp.\ \bibinfo {pages}
  {1--30}\BibitemShut {NoStop}%
\bibitem [{\citenamefont {Curro}(2009)}]{Curro2009}%
  \BibitemOpen
  \bibfield  {author} {\bibinfo {author} {\bibfnamefont {N~J}\ \bibnamefont
  {Curro}},\ }\bibfield  {title} {\enquote {\bibinfo {title} {Nuclear magnetic
  resonance in the heavy fermion superconductors},}\ }\href {\doibase
  10.1088/0034-4885/72/2/026502} {\bibfield  {journal} {\bibinfo  {journal}
  {Rep. Prog. Phys.}\ }\textbf {\bibinfo {volume} {72}},\ \bibinfo {pages}
  {026502 (24pp)} (\bibinfo {year} {2009})}\BibitemShut {NoStop}%
\bibitem [{\citenamefont {Nisson}\ and\ \citenamefont
  {Curro}(2016)}]{NissonCEFSOC2016}%
  \BibitemOpen
  \bibfield  {author} {\bibinfo {author} {\bibfnamefont {D~M}\ \bibnamefont
  {Nisson}}\ and\ \bibinfo {author} {\bibfnamefont {N~J}\ \bibnamefont
  {Curro}},\ }\bibfield  {title} {\enquote {\bibinfo {title} {Nuclear magnetic
  resonance {K}night shifts in the presence of strong spin-orbit and
  crystal-field potentials},}\ }\href
  {http://stacks.iop.org/1367-2630/18/i=7/a=073041} {\bibfield  {journal}
  {\bibinfo  {journal} {New J. Phys.}\ }\textbf {\bibinfo {volume} {18}},\
  \bibinfo {pages} {073041} (\bibinfo {year} {2016})}\BibitemShut {NoStop}%
\bibitem [{\citenamefont {Moriya}(1963)}]{MoriyaT1formula}%
  \BibitemOpen
  \bibfield  {author} {\bibinfo {author} {\bibfnamefont {T\^{o}ru}\
  \bibnamefont {Moriya}},\ }\bibfield  {title} {\enquote {\bibinfo {title} {The
  effect of electron-electron interaction on the nuclear spin relaxation in
  metals},}\ }\href {\doibase 10.1143/JPSJ.18.516} {\bibfield  {journal}
  {\bibinfo  {journal} {J. Phys. Soc. Jpn.}\ }\textbf {\bibinfo {volume}
  {18}},\ \bibinfo {pages} {516--520} (\bibinfo {year} {1963})}\BibitemShut
  {NoStop}%
\bibitem [{\citenamefont {Beeman}\ and\ \citenamefont
  {Pincus}(1968)}]{PhysRev.166.359}%
  \BibitemOpen
  \bibfield  {author} {\bibinfo {author} {\bibfnamefont {D.}~\bibnamefont
  {Beeman}}\ and\ \bibinfo {author} {\bibfnamefont {P.}~\bibnamefont
  {Pincus}},\ }\bibfield  {title} {\enquote {\bibinfo {title} {Nuclear
  spin-lattice relaxation in magnetic insulators},}\ }\href {\doibase
  10.1103/PhysRev.166.359} {\bibfield  {journal} {\bibinfo  {journal} {Phys.
  Rev.}\ }\textbf {\bibinfo {volume} {166}},\ \bibinfo {pages} {359--375}
  (\bibinfo {year} {1968})}\BibitemShut {NoStop}%
\bibitem [{\citenamefont {G\"unther}\ \emph {et~al.}(2014)\citenamefont
  {G\"unther}, \citenamefont {Kamusella}, \citenamefont {Sarkar}, \citenamefont
  {Goltz}, \citenamefont {Luetkens}, \citenamefont {Pascua}, \citenamefont
  {Do}, \citenamefont {Choi}, \citenamefont {Zhou}, \citenamefont {Blum},
  \citenamefont {Wurmehl}, \citenamefont {B\"uchner},\ and\ \citenamefont
  {Klauss}}]{PhysRevB.90.184408}%
  \BibitemOpen
  \bibfield  {author} {\bibinfo {author} {\bibfnamefont {M.}~\bibnamefont
  {G\"unther}}, \bibinfo {author} {\bibfnamefont {S.}~\bibnamefont
  {Kamusella}}, \bibinfo {author} {\bibfnamefont {R.}~\bibnamefont {Sarkar}},
  \bibinfo {author} {\bibfnamefont {T.}~\bibnamefont {Goltz}}, \bibinfo
  {author} {\bibfnamefont {H.}~\bibnamefont {Luetkens}}, \bibinfo {author}
  {\bibfnamefont {G.}~\bibnamefont {Pascua}}, \bibinfo {author} {\bibfnamefont
  {S.-H.}\ \bibnamefont {Do}}, \bibinfo {author} {\bibfnamefont {K.-Y.}\
  \bibnamefont {Choi}}, \bibinfo {author} {\bibfnamefont {H.~D.}\ \bibnamefont
  {Zhou}}, \bibinfo {author} {\bibfnamefont {C.~G.~F.}\ \bibnamefont {Blum}},
  \bibinfo {author} {\bibfnamefont {S.}~\bibnamefont {Wurmehl}}, \bibinfo
  {author} {\bibfnamefont {B.}~\bibnamefont {B\"uchner}}, \ and\ \bibinfo
  {author} {\bibfnamefont {H.-H.}\ \bibnamefont {Klauss}},\ }\bibfield  {title}
  {\enquote {\bibinfo {title} {Magnetic order and spin dynamics in
  ${\mathrm{la}}_{2}{\mathrm{o}}_{2}{\mathrm{fe}}_{2}{\mathrm{ose}}_{2}$ probed
  by $^{57}\mathrm{Fe}$ m\"ossbauer, $^{139}\mathrm{La}$ nmr, and muon-spin
  relaxation spectroscopy},}\ }\href {\doibase 10.1103/PhysRevB.90.184408}
  {\bibfield  {journal} {\bibinfo  {journal} {Phys. Rev. B}\ }\textbf {\bibinfo
  {volume} {90}},\ \bibinfo {pages} {184408} (\bibinfo {year}
  {2014})}\BibitemShut {NoStop}%
\bibitem [{\citenamefont {Knolle}\ \emph {et~al.}(2014)\citenamefont {Knolle},
  \citenamefont {Kovrizhin}, \citenamefont {Chalker},\ and\ \citenamefont
  {Moessner}}]{MoessnerPRL}%
  \BibitemOpen
  \bibfield  {author} {\bibinfo {author} {\bibfnamefont {J.}~\bibnamefont
  {Knolle}}, \bibinfo {author} {\bibfnamefont {D.~L.}\ \bibnamefont
  {Kovrizhin}}, \bibinfo {author} {\bibfnamefont {J.~T.}\ \bibnamefont
  {Chalker}}, \ and\ \bibinfo {author} {\bibfnamefont {R.}~\bibnamefont
  {Moessner}},\ }\bibfield  {title} {\enquote {\bibinfo {title} {Dynamics of a
  two-dimensional quantum spin liquid: Signatures of emergent {M}ajorana
  fermions and fluxes},}\ }\href {\doibase 10.1103/PhysRevLett.112.207203}
  {\bibfield  {journal} {\bibinfo  {journal} {Phys. Rev. Lett.}\ }\textbf
  {\bibinfo {volume} {112}},\ \bibinfo {pages} {207203} (\bibinfo {year}
  {2014})}\BibitemShut {NoStop}%
\bibitem [{\citenamefont {Knolle}\ \emph {et~al.}(2015)\citenamefont {Knolle},
  \citenamefont {Kovrizhin}, \citenamefont {Chalker},\ and\ \citenamefont
  {Moessner}}]{MoessnerPRB}%
  \BibitemOpen
  \bibfield  {author} {\bibinfo {author} {\bibfnamefont {J.}~\bibnamefont
  {Knolle}}, \bibinfo {author} {\bibfnamefont {D.~L.}\ \bibnamefont
  {Kovrizhin}}, \bibinfo {author} {\bibfnamefont {J.~T.}\ \bibnamefont
  {Chalker}}, \ and\ \bibinfo {author} {\bibfnamefont {R.}~\bibnamefont
  {Moessner}},\ }\bibfield  {title} {\enquote {\bibinfo {title} {Dynamics of
  fractionalization in quantum spin liquids},}\ }\href {\doibase
  10.1103/PhysRevB.92.115127} {\bibfield  {journal} {\bibinfo  {journal} {Phys.
  Rev. B}\ }\textbf {\bibinfo {volume} {92}},\ \bibinfo {pages} {115127}
  (\bibinfo {year} {2015})}\BibitemShut {NoStop}%
\bibitem [{\citenamefont {Song}\ \emph {et~al.}(2016)\citenamefont {Song},
  \citenamefont {You},\ and\ \citenamefont {Balents}}]{PhysRevLett.117.037209}%
  \BibitemOpen
  \bibfield  {author} {\bibinfo {author} {\bibfnamefont {Xue-Yang}\
  \bibnamefont {Song}}, \bibinfo {author} {\bibfnamefont {Yi-Zhuang}\
  \bibnamefont {You}}, \ and\ \bibinfo {author} {\bibfnamefont {Leon}\
  \bibnamefont {Balents}},\ }\bibfield  {title} {\enquote {\bibinfo {title}
  {Low-energy spin dynamics of the honeycomb spin liquid beyond the kitaev
  limit},}\ }\href {\doibase 10.1103/PhysRevLett.117.037209} {\bibfield
  {journal} {\bibinfo  {journal} {Phys. Rev. Lett.}\ }\textbf {\bibinfo
  {volume} {117}},\ \bibinfo {pages} {037209} (\bibinfo {year}
  {2016})}\BibitemShut {NoStop}%
\bibitem [{\citenamefont {Khuntia}\ \emph {et~al.}(2017)\citenamefont
  {Khuntia}, \citenamefont {Manni}, \citenamefont {Foronda}, \citenamefont
  {Lancaster}, \citenamefont {Blundell}, \citenamefont {Gegenwart},\ and\
  \citenamefont {Baenitz}}]{PhysRevB.96.094432}%
  \BibitemOpen
  \bibfield  {author} {\bibinfo {author} {\bibfnamefont {P.}~\bibnamefont
  {Khuntia}}, \bibinfo {author} {\bibfnamefont {S.}~\bibnamefont {Manni}},
  \bibinfo {author} {\bibfnamefont {F.~R.}\ \bibnamefont {Foronda}}, \bibinfo
  {author} {\bibfnamefont {T.}~\bibnamefont {Lancaster}}, \bibinfo {author}
  {\bibfnamefont {S.~J.}\ \bibnamefont {Blundell}}, \bibinfo {author}
  {\bibfnamefont {P.}~\bibnamefont {Gegenwart}}, \ and\ \bibinfo {author}
  {\bibfnamefont {M.}~\bibnamefont {Baenitz}},\ }\bibfield  {title} {\enquote
  {\bibinfo {title} {Local magnetism and spin dynamics of the frustrated
  honeycomb rhodate ${\mathrm{li}}_{2}{\mathrm{rho}}_{3}$},}\ }\href {\doibase
  10.1103/PhysRevB.96.094432} {\bibfield  {journal} {\bibinfo  {journal} {Phys.
  Rev. B}\ }\textbf {\bibinfo {volume} {96}},\ \bibinfo {pages} {094432}
  (\bibinfo {year} {2017})}\BibitemShut {NoStop}%
\bibitem [{\citenamefont {Kimchi}\ \emph
  {et~al.}(2018{\natexlab{a}})\citenamefont {Kimchi}, \citenamefont {Nahum},\
  and\ \citenamefont {Senthil}}]{KimchiRandomQuantumMagnets}%
  \BibitemOpen
  \bibfield  {author} {\bibinfo {author} {\bibfnamefont {Itamar}\ \bibnamefont
  {Kimchi}}, \bibinfo {author} {\bibfnamefont {Adam}\ \bibnamefont {Nahum}}, \
  and\ \bibinfo {author} {\bibfnamefont {T.}~\bibnamefont {Senthil}},\
  }\bibfield  {title} {\enquote {\bibinfo {title} {Valence bonds in random
  quantum magnets: Theory and application to ${\mathrm{ybmggao}}_{4}$},}\
  }\href {\doibase 10.1103/PhysRevX.8.031028} {\bibfield  {journal} {\bibinfo
  {journal} {Phys. Rev. X}\ }\textbf {\bibinfo {volume} {8}},\ \bibinfo {pages}
  {031028} (\bibinfo {year} {2018}{\natexlab{a}})}\BibitemShut {NoStop}%
\bibitem [{\citenamefont {Kimchi}\ \emph
  {et~al.}(2018{\natexlab{b}})\citenamefont {Kimchi}, \citenamefont
  {Sheckelton}, \citenamefont {McQueen},\ and\ \citenamefont
  {Lee}}]{Kimchi2018}%
  \BibitemOpen
  \bibfield  {author} {\bibinfo {author} {\bibfnamefont {Itamar}\ \bibnamefont
  {Kimchi}}, \bibinfo {author} {\bibfnamefont {John~P.}\ \bibnamefont
  {Sheckelton}}, \bibinfo {author} {\bibfnamefont {Tyrel~M.}\ \bibnamefont
  {McQueen}}, \ and\ \bibinfo {author} {\bibfnamefont {Patrick~A.}\
  \bibnamefont {Lee}},\ }\bibfield  {title} {\enquote {\bibinfo {title}
  {Scaling and data collapse from local moments in frustrated disordered
  quantum spin systems},}\ }\href {https://doi.org/10.1038/s41467-018-06800-2}
  {\bibfield  {journal} {\bibinfo  {journal} {Nature Communications}\ }\textbf
  {\bibinfo {volume} {9}},\ \bibinfo {pages} {4367} (\bibinfo {year}
  {2018}{\natexlab{b}})}\BibitemShut {NoStop}%
\end{thebibliography}%

\end{document}